\documentclass{IEEEcsmag}

\usepackage[colorlinks,urlcolor=blue,linkcolor=blue,citecolor=blue]{hyperref}
\expandafter\def\expandafter\UrlBreaks\expandafter{\UrlBreaks\do\/\do\*\do\-\do\~\do\'\do\"\do\-}
\usepackage{upmath,color}
\usepackage{booktabs, tabularx}
\usepackage{booktabs, tabularx, xcolor, colortbl}
\usepackage{comment}
\usepackage{ulem}
\usepackage{subcaption}
\usepackage{graphicx}

\definecolor{RowGray}{gray}{0.97}
\definecolor{HeaderGray}{gray}{0.92}

\jvol{XX}
\jnum{XX}
\paper{8}
\jmonth{Month}
\jname{Publication Name}
\jtitle{Publication Title}
\pubyear{2021}

\begin{document}

\sptitle{Article Type: Peer-Reviewed Article \\ \small This work has been submitted to the IEEE for possible publication. Copyright may be transferred without notice, after which this version may no longer be accessible.}

\title{Don’t Guess, Escalate: Towards Explainable Uncertainty-Calibrated AI Forensic Agents}

\author{Giulia Boato}
\affil{Truebees, Trento, 38122, Italy and University of Trento, Trento, 38123, Italy}

\author{Andrea Montibeller}
\affil{Truebees, Trento, 38122, Italy and University of Trento, Trento, 38123, Italy}

\author{Edward Delp}
\affil{Purdue University, West Lafayette, IN 47907, United States}

\author{{L}uisa Verdoliva}
\affil{University of Naples Federico II, Napoli, 80125, Italy}

\author{Daniele Miorandi}
\affil{Truebees, Trento, 38122, Italy}

\markboth{THEME/FEATURE/DEPARTMENT}{THEME/FEATURE/DEPARTMENT}

\begin{abstract}
AI is reshaping the landscape of multimedia forensics.
We propose AI forensic agents: reliable orchestrators that select and combine forensic detectors, identify provenance and context, and provide uncertainty-aware assessments. We highlight pitfalls in current solutions and introduce a unified framework to improve the authenticity verification process.
\\

\end{abstract}

\maketitle

\chapteri{M}ultimedia forensics seeks to determine the authenticity and provenance of digital content such as images, videos, and audio recordings. For many years, the field has relied on handcrafted analytical cues -- noise patterns, compression artifacts, sensor fingerprints, and temporal inconsistencies~\cite{verdoliva2020media}. These methods provide a robust foundation when the acquisition pipeline is well understood. For example, if all images come from a restricted set of camera models and undergo a predictable compression and distribution chain, the resulting data distribution is constrained enough for device-specific artifacts to be modeled and exploited. Likewise, when the threat model is simple -- cut-and-paste forgeries, single-frame insertion, face-swapping -- algorithms can be effectively optimized for that scenario. In such controlled contexts, handcrafted cues remain highly discriminative for distinguishing real from manipulated or synthetic content.

This landscape has changed rapidly with the rise of generative AI. Large vision, language, and audio models can now create highly realistic synthetic images, videos, and voices at scale. Their output challenges long-standing assumptions about what forensic cues remain trustworthy. Moreover, capabilities once limited to expert operators are now widely accessible through user-friendly editing and generation tools. As a result, forensic science must keep pace with an adversarial environment that is evolving quickly and unpredictably.

We argue that the field of multimedia forensics must now shift from the design and implementation of isolated, single-purpose detectors to holistic AI forensic agents: systems capable of autonomously orchestrating multiple tools to detect, characterize, and interpret AI-generated content. This need is driven by two main factors. First, today's ecosystem is fragmented: most detectors are tied to a specific modality or manipulation type, limiting their ability to generalize to emerging or cross-modal threats. Second, in the absence of orchestration, critical information -- e.g., provenance trails, metadata, multimodal correlations, uncertainty estimates -- remains only partially exploited. Existing attempts at combining detectors typically rely on majority voting, weighted aggregation, or simple score-fusion rules. While such ensembles may outperform individual models, their final integrity score is often opaque, offering little insight into which forensic cues influenced the decision.

This position article outlines a vision for trustworthy AI forensic agents. We identify key scientific and technological gaps, from integrating heterogeneous forensic traces to designing explainable approaches that report both an integrity score and a calibrated confidence score -- and that can abstain when evidence is insufficient. We envision AI-agent orchestrators that can enhance explainability by turning low-level forensic outputs from specialized detectors into human-interpretable justifications and coherent narratives about why specific artifacts matter. Our goal is to encourage researchers and practitioners to adopt agent-based orchestration as the next paradigm in multimedia forensics, strengthening the rigor, transparency, and credibility of authenticity assessments in the era of generative AI.

\begin{figure*}[t]
    \centering
    \includegraphics[width=1\linewidth,trim=0 100 0 0, clip]{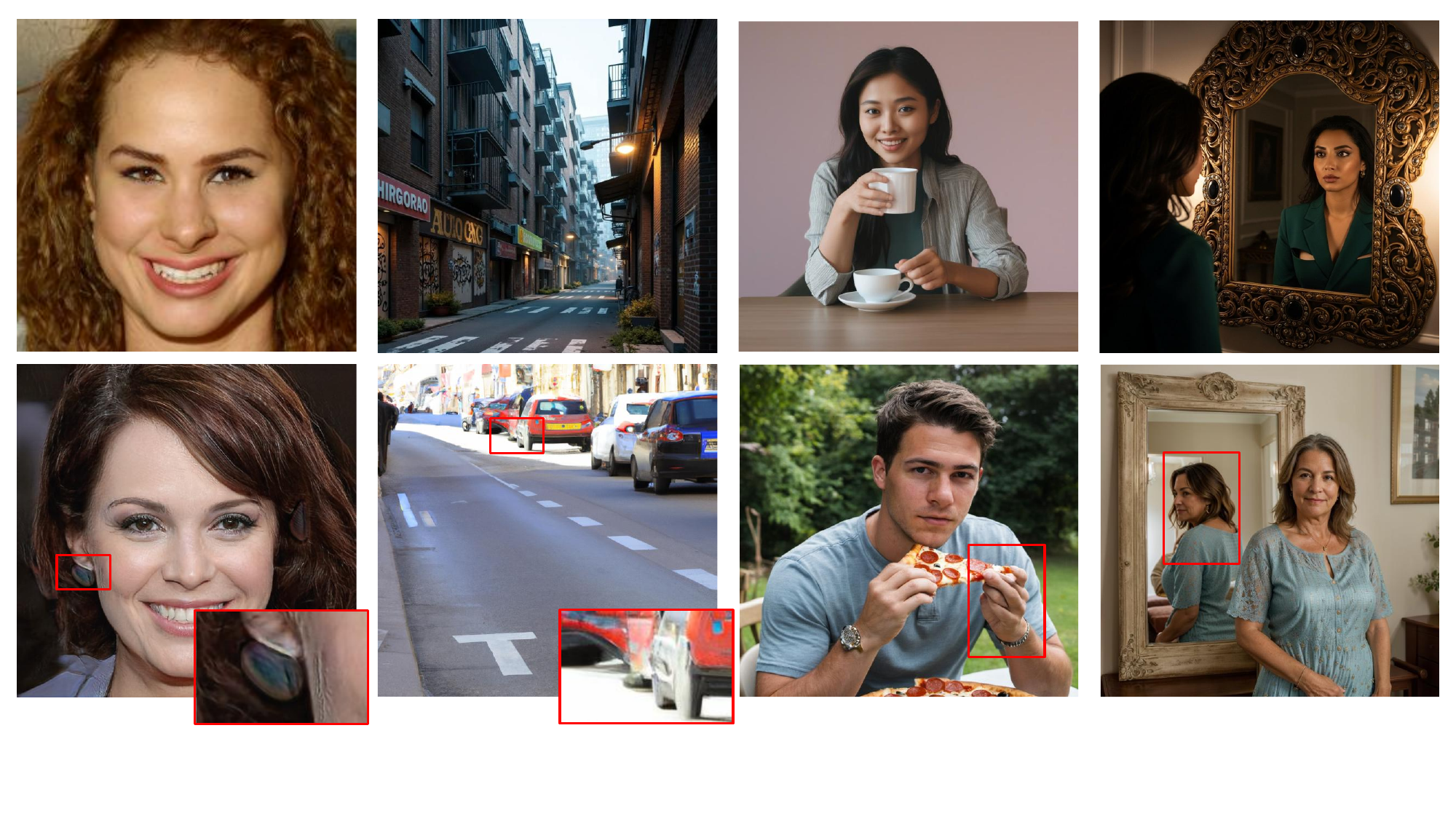}
    \caption{Examples of AI-generated images exhibiting near-photorealistic quality (top) and, conversely, images with obvious visual artifacts (bottom). These cases highlight that some generators leave mainly subtle, low-level traces, while others introduce clear semantic inconsistencies at a high level. Consequently, a detector that relies exclusively on either low-level or high-level artifacts is unlikely to be robust enough to handle the full spectrum of synthetic imagery.}
    \label{fig:examples}
\end{figure*}

\section{AI-GENERATED MEDIA}

Synthetic image generation today is mostly based on two families of models: Generative Adversarial Networks (GANs) and diffusion models (DMs), both capable of high-quality and text-conditioned image synthesis. GANs use an adversarial game between a generator and a discriminator; as they improve against each other, the generator learns to produce increasingly realistic images. Classic milestones include BigGAN and ProGAN, leading to the StyleGAN family, which offers fine control over the synthesis process via latent codes, and EG3D for 3D-consistent face rendering. More recent GANs such as StyleGAN-T, GALIP, and GigaGAN integrate text conditioning and large-scale training, enabling controllable, high-resolution, and fast image generation.

Diffusion models instead learn to invert a noising process: a forward process progressively corrupts real images with Gaussian noise, and a backward process, implemented by neural networks, denoises step by step to synthesize images. They match or improve upon GANs in image quality, are more stable in training, and naturally support text-to-image generation. Many diffusion architectures use a U-Net backbone with a text encoder (e.g., GLIDE, DALL-E 2, eDiff-I), or operate in a latent space to reduce cost (e.g., Latent DM and Stable Diffusion). Variants such as InstructPix2Pix, Stable Diffusion XL, DeepFloyd IF, and Diffusion Transformers (DiTs) push resolution, controllability, and architectural diversity even further \cite{Peebles2023scalable}.

Synthetic video generation has also evolved  very quickly, shifting from early image-based approaches, which often struggled ensuring stable motion and temporal coherence, to specialized video models that more effectively capture space-time structure. Modern architectures using 3D autoencoders and transformers offer better alignment with text prompts and generate higher-quality, more temporally consistent videos. More recently, autoregressive approaches have also achieved impressive video generation performance \cite{deng2025autoregressive}.

In the audio domain \cite{li2025survey}, speech generative techniques have significantly advanced during the last five years, through the implementation of Text-to-Speech and Voice Cloning techniques. The former utilize vocoders to generate natural-sounding speech from text inputs, while the latter modify an original speech signal to mimic a target speaker while preserving the underlying linguistic content. Beyond traditional vocoder-based methods, recent approaches leverage large language models (LLMs) and neural audio codecs to produce highly realistic synthesized voices.

Despite their realism, synthetic images, videos, and audio still carry forensic traces. At a semantic (high) level, early synthetic image and video generators often produced visible artifacts like odd colors, inconsistent lighting, or wrong geometry, but newer models can yield visually flawless images (Fig. \ref{fig:examples}). Nonetheless, every acquisition or generation pipeline leaves characteristic low-level artifacts, analogous to device traces used in traditional image forensics. In generative AI, architecture-specific patterns can be revealed through statistical and frequency-domain analysis, and then exploited for detection or attribution. Common telltale traces are anomalous peaks in the power spectra of noise residuals and spectral distributions that do not match the real data in the mid- to high-frequency range.
Just as with visual media, audio generation algorithms -- ranging from WaveNet to modern neural codecs -- leave distinct forensic traces, such as vocoder fingerprints and spectral-magnitude anomalies, that can be used to distinguish synthetic speech from bona fide human audio. 

\section{THE SHIFTING LANDSCAPE OF MULTIMEDIA FORENSICS}

Classic methods exploit artifacts introduced either by the in-camera processing chain or by subsequent editing operations, 
which leave distinctive traces on the acquired image \cite{verdoliva2020media}.
For instance, different camera models usually adopt different demosaicing algorithms, so when a forgery combines regions coming from multiple devices, inconsistencies in the demosaicing pattern may appear. Similarly, out-of-camera editing can leave its own characteristic marks and may also alter or suppress sensor-specific fingerprint patterns, both of which can be exploited to reveal tampering. Exploiting compression artifacts has long been a cornerstone of image forensics, for example by leveraging the block artifact grid or double-compression traces. A more general strategy is to analyze noise artifacts introduced by the entire camera-based acquisition process, for example by examining local noise levels or high-pass noise residuals to expose inconsistencies between regions. Although these traces are extremely subtle and invisible to the naked eye, once suitably enhanced and analyzed they provide a valuable source of evidence for assessing the integrity of digital content. 

These methods are powerful but rely heavily on specific statistical models, and their performance worsens when the underlying assumptions are not satisfied. For example, they are very sensitive to changes in resolution or to additional compression, both of which are very common on social media platforms \cite{diangarti2024synthetic}. In such challenging conditions, deep learning, especially very deep architectures, can offer a substantial performance gain over conventional methods. However, the effectiveness of data-driven
approaches may be overemphasized due to poorly designed training and testing protocols. Indeed, these methods perform well when training and test data are perfectly aligned, but suffer a significant drop in accuracy under distribution shifts. Detecting images generated by known generative models is relatively straightforward; the real challenge is to generalize to synthetic content produced by unseen generators. This issue is particularly pressing today, as new generative models are being released at an almost continuous pace.
In this context, large pre-trained vision–language models (VLMs) such as CLIP have shown remarkable robustness to distribution shifts. Their success in forensic applications suggests that pre-training on large and diverse datasets can foster more semantic, model-agnostic representations, enabling detectors that generalize better to unseen manipulations and generative models, even when no examples from these sources are available at training time. 

Furthermore, recent work is increasingly leveraging multimodal large models to obtain more transparent and interpretable forensic decisions \cite{zou2025survey}. By means of tailored prompt learning, these approaches can generate human-readable explanations that explicitly relate subtle visual artifacts and inconsistencies to the final classification outcome, thereby making the decision process easier to inspect \cite{Huang2025sida, Yu2025unlocking}. 
However, although such tools can generate coherent explanations and detect semantic anomalies, they are unable to reason jointly over multiple sources of forensic evidence or to improve over time through experience. They operate almost exclusively on image content and ignore other critical forensic information, such as metadata, provenance information, and contextual cues. As a result, they function as standalone detectors with highly limited capacity for context-aware reasoning.   
   
\begin{figure*}[t]
\centerline{\includegraphics[width=\textwidth]{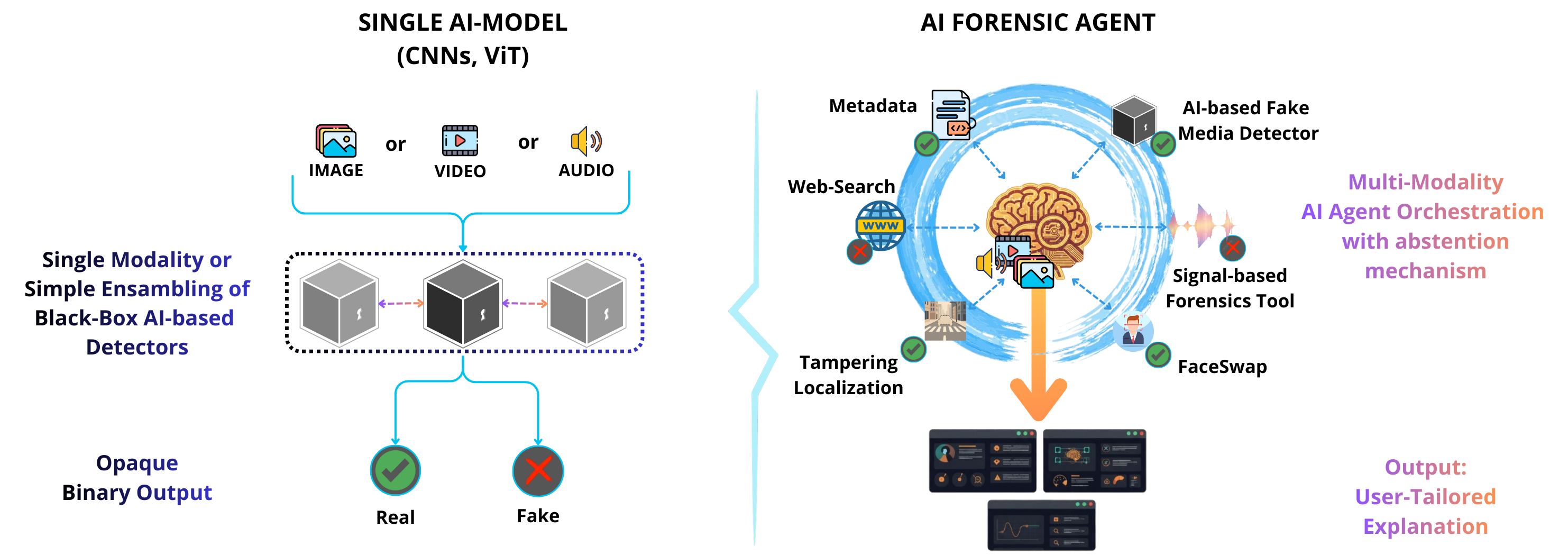}}
\caption{
Re-imagining multimedia forensics: transitioning from single-detector systems to AI forensic orchestrators. Left: traditional ensembles combining a fixed set of models into a single, often opaque confidence score. Right: the AI forensic agent paradigm, where diverse tools -- e.g., low-level forensic traces, face biometrics, metadata analysis, and web-based provenance retrieval -- are dynamically selected, analyzed, and fused. This agent-based architecture enables explicit reasoning, calibrated uncertainty, and adaptive integration of new detectors, resulting in more robust and transparent forensic analysis.
}\vspace*{-5pt}
\label{fig:single_v_orchestrator}
\end{figure*}

\section{FROM SINGLE DETECTORS TO ORCHESTRATORS}
Current multimedia forensic detectors -- whether designed to spot AI-generated media, image inpainting, video manipulation, or voice cloning -- are often treated as {\it isolated sources of truth}. Each method is typically engineered for specific manipulations or forensic traces and is benchmarked in controlled experimental setups. This tends to produce systems that \textbf{lack generalization}: a detector that performs well today can fail tomorrow as generative techniques evolve, datasets drift, or adversaries exploit blind spots. Over-reliance on any single model creates a dangerous asymmetry where attackers evolve their strategies while defenders leverage one single tool.

However, modern manipulations are no longer single edits but pipelines of transformations that leave overlapping and sometimes conflicting traces. When the sequence of operations is unknown -- as is common in real-world investigations -- analysis should shift from isolated detectors to unified reasoning systems. The objective is to synthesize heterogeneous cues into a coherent, uncertainty-aware assessment of authenticity. 

Recent research has explored ensembles of detectors, combining multiple models through majority voting, weighted aggregation, or related fusion rules~\cite{SILVA2022100217}. These ensembles could outperform individual models by exploiting complementary features, but they inherit a significant limitation: {\it opacity}. The final confidence score is a black-box combination of the individual detectors' specific features and the ad hoc fusion heuristics, potentially encoding non-trivial biases that are difficult to anticipate or interpret. As a result, ensemble systems often struggle to generalize to unseen manipulation types or emerging generative models, while also increasing computational overhead.

These limitations indicate the need for a deeper conceptual reframing. Instead of simply adding more detectors, multimedia forensics should transition toward {\bf AI forensic orchestrators} -- systems that do not only fuse outputs but actively reason over a diverse set of forensic tools and specialised sub‑agents within a unified framework  (Fig.~\ref{fig:single_v_orchestrator}). Similar orchestration concepts \cite{10.1145/3701716.3715232,peigne2025multi} have been adopted in domains such as automated trading and financial analytics, where agents dynamically manage portfolios of specialized models to achieve complex objectives. In a forensic context, an orchestrator plays a comparable role: it decides which tools or sub‑agents to invoke, in what order, and how their outputs should be interpreted. 

A forensic orchestrator can activate the right component at the right time, assess each detector’s reliability in light of contextual information, and calibrate uncertainty across heterogeneous cues. Its scope extends beyond aggregation: it can also abstain when evidence is weak or contradictory, transforming uncertainty from a failure mode into a safeguard against overconfident conclusions. 
Orchestration also enables richer reasoning. By integrating cross‑modal information, coordinating provenance reconstruction across multiple transformations, and tailoring explanations to different user profiles (e.g., investigators, journalists, or legal decision-makers), orchestrators provide a level of interpretability that simple detector fusion cannot achieve. In this way, low‑level traces, semantic inconsistencies, metadata and contextual cues can be combined into a coherent, evidence‑grounded assessment.

\begin{table*}[ht]
\centering
\caption{From standalone detectors to orchestrators: contrasting today’s multimedia forensic practice with the emerging paradigm of AI forensic agents.}
\vspace{4pt}
\renewcommand{\arraystretch}{1.25}
\setlength{\tabcolsep}{8pt}

\begin{tabularx}{\textwidth}{@{}p{3.2cm}X X@{}}

\toprule
\textbf{Aspect} &
\textbf{Today’s Practice} &
\textbf{Proposed Vision} \\
\midrule
\rowcolor{RowGray}
\textbf{Role of Detectors} &
Detectors operate as isolated sources of truth, each focused on a specific manipulation or modality. & Specialized tools orchestrated dynamically based on context, available evidence, and investigative goals.\\


\textbf{Fusion of Results} &
Weak or ad-hoc fusion (e.g., manual thresholds, majority voting), often lacking principled uncertainty handling. & Probabilistic reasoning and structured fusion across modalities, incorporating uncertainty, reliability, and complementary signals.\\

\rowcolor{RowGray}
\textbf{Adaptability} &
Fragile to new generative techniques; high risk of rapid obsolescence. & Modular, continuously adaptive orchestration informed by prior cases and evolving manipulation patterns.\\

\textbf{Provenance} &
Limited visibility; training data, decision paths, and tool interactions rarely traceable. & Full provenance trails detailing which tools were used, in which sequence, and how each contributed to the final assessment.\\

\rowcolor{RowGray}
\textbf{Uncertainty Handling} &
Binary decisions (real/fake) or poorly calibrated scores; little transparency about confidence. & 
Calibrated probabilities with principled abstention when evidence is insufficient or contradictory.
\\

\textbf{User Interaction} &
One-size-fits-all outputs that can be opaque to non-experts. & Explanations tailored to user expertise (e.g., investigator, journalist, legal professional), with clear rationale.
\\

\rowcolor{RowGray}
\textbf{Scope} &
Single-modality focus or reliance on standalone detectors. & Multimodal orchestration integrating image, video, audio, metadata, and cross-modal reasoning. \\
\bottomrule
\end{tabularx}

\vspace{6pt}
\label{tab:ai_forensics_comparison}
\end{table*}

\section{TRUSTWORTHY FORENSIC AGENTS}

\begin{figure}
\centerline{\includegraphics[width=\linewidth]{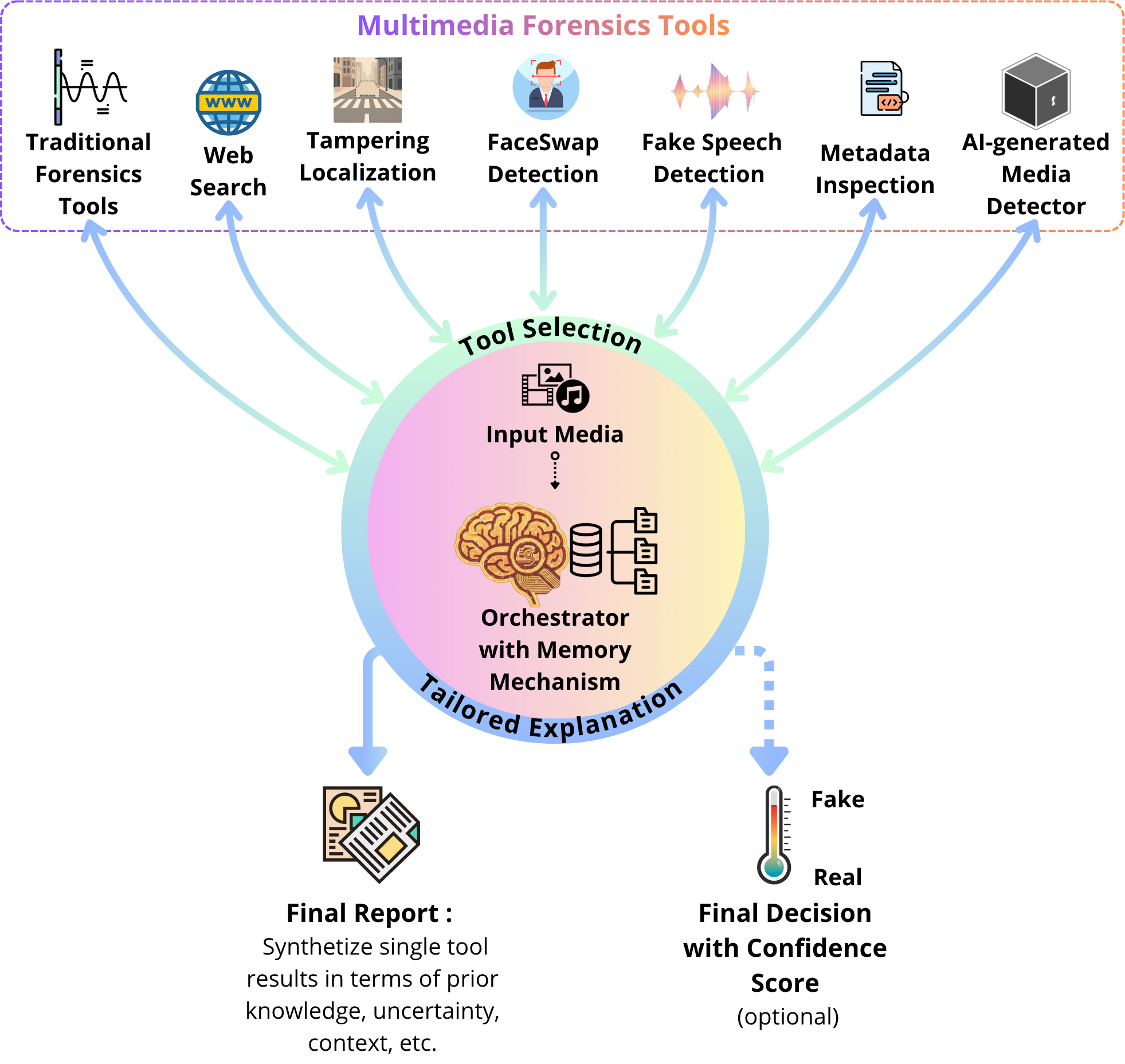}}
\caption{Conceptual architecture of an AI forensic agent. The central orchestrator selects and combines multiple detection tools -- ranging from low-level forensic analyses to metadata and provenance checks -- and synthesizes their outputs into a transparent final report with calibrated confidence scores.}\vspace*{-5pt}
\label{fig:orchestrator}
\end{figure}

In the proposed vision, a multimedia forensics system is a coordinated framework of specialized tools. At its core (Fig.~\ref{fig:orchestrator}), an AI-driven orchestrator determines which tools to use, in what order, and for what purpose, conditioned on the characteristics of the media under analysis.
When a piece of content is submitted, the orchestrator first analyzes its semantics and technical attributes, such as metadata, format, compression history, and any available online related content. Using web-search or provenance analysis (e.g., \`a la C2PA\footnote{https://c2pa.org/ C2PA is an open standard for publishers, creators and consumers to establish the origin and edits of digital content.}), it can reconstruct the media’s life cycle before conducting the analysis with targeted forensic tools. These may include biometric consistency checks, perspective/shadows analysis, fingerprints extraction and/or some more general algorithms for synthetic-versus-real classification and tampering localization. The orchestrator then synthesizes this evidence into a structured, transparent report that explains the reasoning behind its conclusions, including why certain cues were weighted more heavily (e.g., detector training domain, reliability history, modality consistency).

Given that forensic tools expose heterogeneous outputs, such as detection, similarity and anomaly scores, then the AI-agent should properly fuse these indicators: central to this process is a rigorous {\bf quantification of uncertainty}. The system should express not only what it predicts but {\it how confident it is}, and whether it should abstain when evidence is insufficient.
In current multimedia forensics practice, uncertainty is often treated superficially: confidence scores are rarely calibrated, and their behavior under distribution shifts is largely unknown. 

Adjacent safety‑critical domains -- such as medical imaging, weather forecasting, financial risk modeling, and autonomous driving -- offer more mature approaches the community could build upon. Techniques like temperature scaling, deep ensembles, Bayesian neural networks, and proper scoring rules demonstrate how probabilistic predictions can be tied to empirical error rates and systematically evaluated for calibration and discrimination \cite{Rahaman2021uncertainty}.
When detectors disagree or uncertainty is high, the forensic agent should adopt {\bf selective prediction} or {\bf reject‑option} mechanisms, abstaining rather than producing a misleading claim. Frameworks such as {\it conformal prediction} formalize this behavior by guaranteeing error bounds on retained cases. These approaches consistently reduce overconfidence errors and improve user trust -- properties equally crucial in forensic analysis, where false alarms can damage reputations or influence journalistic and legal outcomes.

A further challenge lies in aggregating uncertainty across detectors operating on different cues: low‑level fingerprints, semantic inconsistencies, metadata traces, or cross‑modal correlations. Fusion should consider each tool's reliability, account for correlations, and penalize overconfidence errors. Probabilistic graphical models, ensemble Bayesian inference, and uncertainty‑aware multimodal fusion provide promising directions. Proper scoring rules such as the Brier score, although not yet standard in multimedia forensics, offer a principled way to evaluate how well aggregated probabilities reflect real‑world uncertainty.


Contextual awareness is also crucial for making reliable decisions. Different real-world settings, such as newsrooms, social media platforms or law enforcement investigations, require different evaluation metrics and different ways of weighting evidence, and different decision thresholds.
To support this, the orchestrator can use memory mechanisms that store information about prior cases, model behaviors, and domain-specific constraints. This allows the system to adapt its behaviour and to avoid repeating known failure modes (for example, errors that have previously led to false alarms or missed detections).

From an accountability standpoint, the system should provide a clear record of how each piece of content was analyzed: 
which models and versions were used, with which parameters, in which sequence, and by which analyst or system component. 
This supports a robust “digital chain of custody” that allows to reproduce the verification process. 
The framework should also make it possible to explain methods, uncertainty, and limitations in a way that non-technical stakeholders (judges, lawyers, journalists) can understand, and it should allow results to updated as better tools or new evidence become available.

Table \ref{tab:ai_forensics_comparison} summarizes the proposed vision. Taken together, these principles chart a path toward multimedia forensic agents that are both technically powerful and operationally responsible. With appropriate governance, orchestrated systems can enhance investigative and verification workflows while preserving the human authority, transparency, and ethical safeguards.

\section{TOWARDS EXPLAINABILITY}

{\bf Explainability} is essential for building trust in multimedia forensic agents. These systems have to provide transparent and interpretable outputs that help users understand not only what conclusion has been reached, but why that conclusion is warranted. Researchers, engineers, journalists, and platform operators should be able to follow the reasoning behind an assessment and judge whether the supporting evidence is sufficient. Trust is earned through clarity, reproducibility, and a consistent link between a system’s inferences and the underlying forensic cues.

\subsection{Current approaches and their limitations}

Recent advances in explainable AI (XAI) offer useful principles, but applying them to multimedia forensics poses significant challenges. In computer vision fields such as object recognition, saliency maps and concept-based explanations help articulate a model’s reasoning. However, forensic traces are often subtle, intentionally obfuscated and can involve multiple modalities. This makes difficult to apply existing XAI methods to explanations related to noise-level inconsistencies, compression artifacts, metadata anomalies, and cross-modal correlations, all of which may be crucial to a forensic assessment.

\begin{figure}
    \centering
    \includegraphics[width=1\linewidth]{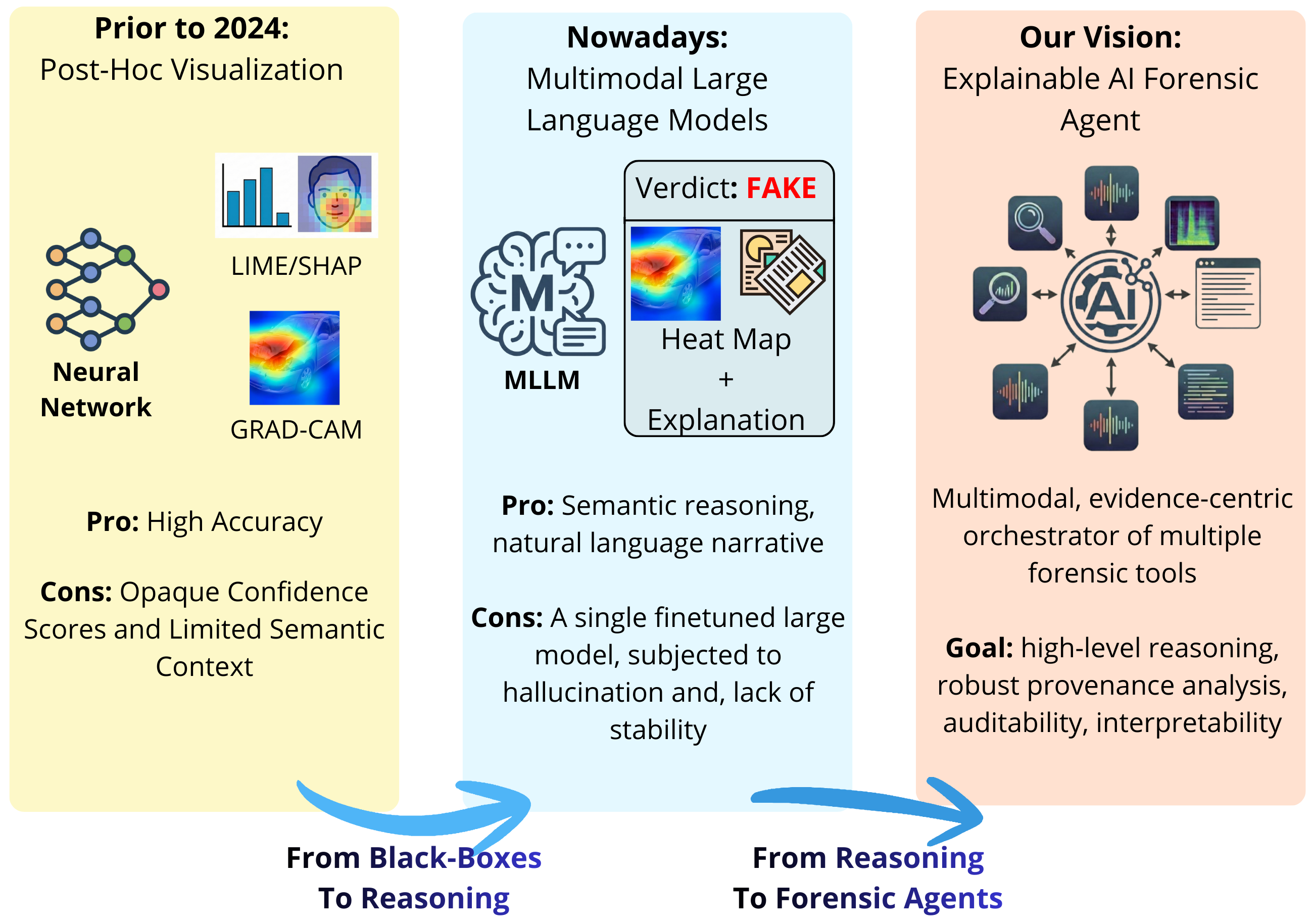}
        \caption{
        Evolution of explainable AI in digital forensics. Left: early ``black-box'' neural networks relying on post-hoc visualizations. Center: emerging semantic reasoning capabilities enabled by modern multimodal large language models (MLLMs). Right: the proposed explainable AI forensic agent, which integrates evidence-focused tools to support improved reasoning, auditability, and interpretability.
        }
    \label{fig:xai}
\end{figure}

The trajectory of XAI in multimedia forensics is still in its infancy, as depicted in Fig.~\ref{fig:xai}. Before 2024, the field was predominantly focused on post-hoc visualization techniques used to highlight the decision-making processes of deep learning models, by exploiting computer vision methods such as Gradient-weighted Class Activation Mapping (Grad-CAM), LIME, and SHAP generated heat maps \cite{momin2025explainable}. Although these techniques can identify visual artifacts, such as blending boundaries or color anomalies, they
are not able to provide clear evidence when forensic traces are not localized but distributed across the entire media, such as in the case of fully AI-generated content. In addition, post-hoc visualizations obtained with different methods often highlight different regions, leading to contradictory indications that are difficult for non-expert users to interpret. Moreover, approaches based on physical or physiological signals (e.g., specular corneal highlights or irregular pupil shapes) are constrained by strict environmental assumptions, limiting their applicability in unconstrained settings \cite{verdoliva2020media}.

Multimodal large language models (MLLMs) have recently generated considerable interest as potential components for explainable forensics, but their current capabilities remain insufficient for reliable use in this domain. Emerging approaches explore finetuning MLLMs to perform forensic tasks ranging from binary detection to fine-grained forgery localization, thereby opening a path toward more advanced semantic reasoning \cite{li2025fakebench}. While they can produce high-level descriptions and flag coarse visual inconsistencies, this analysis is still limited to semantic flaws.


\subsection{Our vision of explainable forensic AI agents}
To move beyond these limitations, future explainability research in multimedia forensics should embrace hybrid, evidence-centric approaches that explicitly link model decisions to measurable, domain-specific forensic cues. Rather than relying solely on post-hoc visualizations, next-generation forensic agents should indicate which signal was detected, how it was measured, and why it is indicative of a manipulation. Explanations grounded in quantifiable artifacts -- frequency inconsistencies, sensor-pattern perturbations, illumination geometry, motion-field deviations, or metadata–pixel mismatches -- are more likely to withstand adversarial conditions and support human verification.

In our proposed vision of AI-agent orchestrators, MLLMs may play a constructive role allowing explainability through human interpretable justifications.
They can organize the outputs of specialized detectors into coherent narratives and help users interpret why certain artifacts matter by translating low-level forensic evidence into \textbf{higher-level reasoning}. 
In this respect, chain-of-thought reasoning can be used to obtain explanations that explicitly connect detector outputs to the final decision. This allows to make clear how each cue contributes to the final decision and highlights when such a decision is driven by specific types of evidence or limited by ambiguity. Within an AI-agent orchestrator, such chain-of-thought style justifications serve as an interface layer between low-level forensic analysis and human decision-makers, turning raw integrity scores into transparent reasoning analysis. A further step would be to consider a multi-agent settings, in which each agent independently proposes answers and then jointly debates their responses and reasoning to reach a shared conclusion. This iterative interaction encourages to construct answers that improve factual validity and reduce hallucinations \cite{du2024improving}.

Explainability also benefits from \textbf{robust provenance analysis}. Understanding where a piece of media originates, how it has been transformed, and through which platforms it has circulated provides context that strengthens the interpretability of authenticity assessments. Combining passive forensic techniques with active methods such as watermarking or C2PA-based provenance can help clarify whether observed traces stem from benign editing workflows or not.  


Finally, explainability should coexist with \textbf{privacy protection}. Multimedia verification frequently involves sensitive material, and explanations should not disclose unnecessary personal details. Secure processing, restricted access, and the use of feature-level representations where appropriate help ensure that the trust gained through transparency is not undermined by privacy risks.

\section{CONCLUSION}

All civilized societies rely on mechanisms of trust. Generative AI can be easily exploited by malicious actors to undermine \textbf{social trust} by spreading misinformation. The recent explosion of AI-generated media risks fostering a general belief that what we see or hear is likely to be synthetic or fake. This, in turn, can erode social cohesion and compromise democratic processes.
To mitigate this problem, we propose AI forensic orchestrators that can provide explainable and reliable tools. To this end, the following research challenges should be addressed: 
\begin{itemize}
    \item Uncertainty-based analysis: forensic agents that orchestrate diverse detectors across modalities and provide calibrated confidence scores and explicitly refrain from deciding when uncertainty is high, thus adhering to the principle: \textbf{don’t guess, escalate}.
    \item Provenance tracking: forensic agents that combine active and passive approaches like C2PA, digital watermarking of generative AI, synthetic versus real forensic detectors. This framework has to address highly challenging scenarios, including tracing sharing chains across social media platforms, where platform-specific compression, resizing, and filtering weaken forensic traces.
    \item Explainability: forensic agents that can provide interpretable results, both in terms of low- and high-level forensic traces. This would ensure legal admissibility of our framework, allowing integration into investigative and judicial workflows.  
\end{itemize}
Overall, we propose the adoption of AI forensic agents that are able to combine a diverse set of tools, from passive to active methods, from detection to attribution, and reason over heterogeneous evidence. The proposed framework is required to provide calibrated confidence scores and, when the available cues are weak or conflicting, to avoid making a definitive decision, thereby reducing the risk of overconfident predictions. This demands domain-aware reasoning to provide interpretable and scientifically grounded justifications. Only through such integrated and transparent forms of reasoning forensic agents can achieve the level of trust necessary for deployment in real-world operational settings.




\section{ACKNOWLEDGMENTS}
This work has been funded by the European Union under Horizon Europe Grant Agreement no. 101092887 (NGI Sargasso -- Open Call OC5 | Deepshield Project).

\def\refname{REFERENCES}
\bibliographystyle{IEEEtran}
\bibliography{references}

\begin{IEEEbiography}{Giulia Boato}{\,} is Full Professor at the University of Trento (Italy) and Co-Founder of Truebees. Since 2012 she is leading the group within the Media Lab working on multimedia forensics. Her research interests include image and signal processing, with particular attention to multimedia data protection and authentication, data hiding and digital forensics, but also intelligent multidimensional data management and analysis. She has published more than 170 papers in international conferences and journals. Contact her at giulia.boato@unitn.it.
\end{IEEEbiography}

\begin{IEEEbiography}{Andrea Montibeller}{\,} is Co-Founder \& Chief AI Officer of Truebees srl, and Assistant Professor at the University of Trento, Italy. He received his Ph.D. in ICT from the University of Trento in 2024, following an M.Sc. in 2020. He has held research scholar positions at Drexel University (2023) and the University of Vigo (2018). His research interests include AI-generated image/video forgery detection and localization, and camera source attribution. Contact him at andrea@truebees.eu. 
\end{IEEEbiography}

\begin{IEEEbiography}{Edward J. Delp}{\,} was born in Cincinnati, Ohio.  He is The Charles William Harrison Distinguished Professor of Electrical and Computer Engineering and Professor of Biomedical Engineering at Purdue University in the United States. His research interests include image analysis, computer vision, machine learning, image and video compression, multimedia security, medical imaging, multimedia systems, communication, and information theory. Dr. Delp is a Life Fellow of the IEEE, a Fellow of the ACM, a Fellow of the SPIE, a Fellow of the Society for Imaging Science and Technology (IS\&T), Fellow of the American Institute of Medical and Biological Engineering,
and a Fellow of the National Academy of Inventors.
Contact him at ace@purdue.edu.
\end{IEEEbiography}

\begin{IEEEbiography}{Luisa Verdoliva} {\,} is Full Professor with the Department of
Electrical Engineering and Information Technology
at University Federico II of Naples, 80125 Naples,
Italy. Her research interests are in the field of image
and video processing, with main contributions in the
area of multimedia forensics. Verdoliva received a
Ph.D. in Information Engineering from the University
Federico II of Naples. She is Editor-in-chief for
IEEE Transactions on Information Forensics and Security and the recipient of the 2025 Frontiers of Science Award, the 2018 Google Faculty 
Award and the TUM Institute
for Advanced Study Hans Fischer Senior Fellowship.
She is a Fellow of IEEE. Contact her at verdoliv@unina.it.
\end{IEEEbiography}

\begin{IEEEbiography}{Daniele Miorandi}{\,} is the CEO of Truebees and Head of AI at Afliant (formerly U-Hopper), and co-founder of Trentino AI. A serial entrepreneur, his current research and business interests include trustworthy AI, multimedia forensics, and privacy-preserving analytics. He has published around 200 scientific articles and papers on IoT, big data platforms, and applied AI. Miorandi received a Ph.D. in Communications Engineering from the University of Padova, Italy. Contact him at daniele@truebees.eu.
\end{IEEEbiography}

\end{document}